\documentclass[11pt]{article}
\usepackage{times}
\usepackage{amsmath,amsfonts,amssymb,latexsym,epsfig,a4}
\usepackage{color,epsf}
\usepackage{graphicx}

\usepackage{alltt}
\usepackage{url}

\newtheorem{proposition}{Proposition}

\newcommand{\e}{\ensuremath{\mathrm{e}}}

\begin{document}
\title{A note on the Baker--Campbell--Hausdorff series in terms of right-nested commutators
}

\author{Ana Arnal\footnote{ Email: \texttt{ana.arnal@uji.es}. ORCD: 0000-0002-3283-3379}
 \and
 Fernando Casas\footnote{Email: \texttt{Fernando.Casas@uji.es}. ORCID: 0000-0002-6445-279X} 
 \and Cristina Chiralt\footnote{Email: \texttt{chiralt@uji.es}. ORCID: 0000-0003-0925-4034}
}

\date{}

\maketitle

\begin{center}
Institut de Matem\`atiques i Aplicacions de Castell\'o (IMAC) and  De\-par\-ta\-ment de
Ma\-te\-m\`a\-ti\-ques, Universitat Jaume I,
  E-12071 Cas\-te\-ll\'on, Spain.
\end{center}

\vspace*{0.5cm}

\begin{abstract}

We get compact expressions for the Baker--Campbell--Hausdorff series $Z = \log(\e^X \, \e^Y)$ in terms of right-nested commutators. The reduction in the number of terms
originates from two facts: (i) we use as a starting point an explicit expression directly involving independent commutators and (ii) we derive
a complete set of identities arising among right-nested commutators.  The procedure allows us to obtain the series with fewer terms than when expressed in the classical
Hall basis at least up to terms of grade 10.

\



\end{abstract}\bigskip



\section{Introduction}

Exponentials of non-commuting operators appear in many areas of physics and mathematics, ranging from quantum mechanics to 
the theory of Lie groups and Lie algebras to the numerical analysis of differential equations. It is then natural to consider products
of such exponentials and how to express such products as the exponential of a new operator. This of course is closely related with the celebrated Baker--Campbell--Hausdorff
theorem \cite{bonfiglioli12tin}. 

In the most basic algebraic setting,
one considers the associative algebra $\mathbb{K}\langle X, Y \rangle$ of formal power series in  the non-commuting variables
$X$ and $Y$ over a field $\mathbb{K}$ of characteristic zero. Then  $\e^X \, \e^Y = \e^{\Phi(X,Y)}$, with
\begin{equation}  \label{zwc}
\aligned
   \Phi(X,Y)  & = \log( \e^X \, \e^Y) = \sum_{k \ge 1} \frac{(-1)^{k-1}}{k} (\e^{X} \e^{Y} -1)^k = \sum_{k \ge 1} \frac{(-1)^{k-1}}{k}
   \left( \sum_{p+q > 0} \frac{X^p Y^q}{p! q!} \right)^k \\   
   & =  \sum_{k \ge1} \frac{(-1)^{k-1}}{k}
   \sum \frac{X^{p_1} Y^{q_1} \ldots X^{p_k} Y^{q_k}}{p_1! \,
      q_1! \ldots p_k! \, q_k!},
\endaligned      
\end{equation}
where, in the last expression, the inner summation extends over all non-negative integers
$p_1$, $q_1$, \ldots, $p_k$, $q_k$ for which $p_i + q_i > 0$
($i=1,2, \ldots, k$). The first terms read explicitly
 \[
 \aligned
  & \Phi =  (X + Y + X Y + \frac{1}{2} X^2 + \frac{1}{2} Y^2 + \cdots )   
 - \frac{1}{2} ( XY + YX + X^2 + Y^2 + \cdots) + \cdots \\
  & =  X + Y + \frac{1}{2} (X Y - Y X) + \cdots = X + Y + \frac{1}{2} [X,Y] + \cdots
\endaligned 
 \]
The Baker--Campbell--Hausdorff (BCH) theorem states that $\Phi(X,Y)$ in (\ref{zwc}) can be expressed
as 
 \begin{equation}  \label{eq.1.3}
   \Phi(X,Y)  = X + Y +
    \sum_{m \ge 2} \, \Phi_m(X,Y),
\end{equation}
where $\Phi_m(X,Y)$ is a \emph{homogeneous Lie polynomial} in $X$ and $Y$
of degree $m$, i.e., a linear combination
of commutators of the form $[V_1,[V_2, \ldots,[V_{m-1},V_m]
\ldots]]$ with $V_i \in \{X,Y\}$ for $1 \le i \le m$, the
coefficients being rational constants. The formal power series (\ref{eq.1.3}) is called the Baker--Campbell--Hausdorff series, and 
plays a fundamental role not only in the theory of Lie groups and Lie algebras \cite{bonfiglioli12tin}, but also in linear differential equations, control theory,
quantum and statistical mechanics, and numerical analysis (see e.g. \cite{bonfiglioli10aov,sornborger99hom,strichartz87tcb,varadarajan84lgl,wilcox67eoa}).

An explicit expression for $\Phi_n$ in the BCH series was provided
 by Dynkin \cite{dynkin47eot,dynkin00spo} in the form
\begin{equation}  \label{eq.1.2}
  \Phi_m(X,Y) =  \sum_{p_i, q_i}
    \frac{(-1)^{m-1}}{m} \frac{[X^{p_1} Y^{q_1} \ldots
      X^{p_m} Y^{q_m} ]}{(\sum_{i=1}^m (p_i + q_i)) \,
      p_1! \, q_1! \ldots p_m! \, q_m!},
\end{equation} 
where the summation is taken over all non-negative
integers $p_1, q_1, \ldots$, $p_m$, $q_m$ such that
$p_1+q_1>0, \ldots, p_m + q_m > 0$ and 
$[ X^{p_1} Y^{q_1} \ldots X^{p_m} Y^{q_m}]$ denotes the right-nested commutator based on the \emph{word}
$ X^{p_1} Y^{q_1} \ldots X^{p_m} Y^{q_m}$, i.e., 
\[
   [XY^2X^2 Y] \equiv [X Y Y X X Y] \equiv [X,[Y,[Y,[X,[X,Y]]]]].
\]
Expression (\ref{eq.1.2}) can be used in principle to compute $\Phi_n$ in the BCH series  up to
any desired order. One should notice, however, is that not all the terms are independent, due to 
the many existing redundancies. Thus, for instance,
$[X^3 Y^1] = [X^1 Y^0 X^2 Y^1] = [X,[X,[X,Y]]]$. An additional source of redundancies arises from the Jacobi identity \cite{varadarajan84lgl}:
\begin{equation}  \label{jacobi}
    [X_1,[X_2,X_3]] + [X_2,[X_3,X_1]] + [X_3,[X_1,X_2]] = 0,
\end{equation}    
for any three variables $X_1, X_2, X_3$, and other identities obtained from it. From this perspective, a procedure allowing to remove at once all the superfluous
terms in (\ref{eq.1.2}) would be of great value for practical applications.

Although different procedures exist in the literature to construct the BCH series up to an arbitrary degree in terms of commutators,
all of them have a basic limitation, as is the case with the Dynkin presentation (\ref{eq.1.2}): not all the commutators are 
independent, and so a rewriting process has to be
carried out to express the results in terms of a basis of the free Lie algebra $\mathcal{L}(X,Y)$ generated by $X$ and $Y$. This process, of course, although can be
carried out by computer algebra systems, 
requires a good deal of
computational time and memory resources.
One of the most efficient
algorithms was proposed in \cite{casas09aea}, where explicit expressions of $\Phi_{m}$ up to $m=20$
in terms of the classical Hall and Lyndon basis of $\mathcal{L}(X,Y)$ were obtained with relatively modest computer requirements. 
In any event, the fact that no basis in the free Lie algebra
exists that eases the calculation of the BCH series is one major problem when dealing with problems where this series 
plays a role \cite{burgunder08eia}.

Expressing the BCH series in terms of right-nested commutators presents several advantages, especially when the series is
considered in some particular physical settings. There are problems whose structure leads in a natural way to consider the
BCH series of two operators $X$, $Y$ satisfying $[Y,[Y,[X,Y]]] \equiv 0$. This happens, in particular, when designing 
splitting methods for the numerical integration of classical Hamiltonian systems and also for the time-dependent Schr\"odinger
equation. Although in this case it is still possible to construct a generalized Hall basis \cite{mclachlan19tla}, it is much simpler
to identify the non-vanishing terms when using right-nested commutators. On the other hand, and contrary to Hall--Viennot
bases, there is no a straightforward procedure to construct a set of independent right-nested commutators generating each
homogeneous subspace of $\mathcal{L}(X,Y)$.

Several attempts have been made to directly remove in (\ref{eq.1.2}) redundant terms and therefore to express $\Phi_m$ only as
a linear combination of independent right-nested commutators. Thus, we can mention in particular references \cite{oteo91tbc}
and \cite{kolsrud93mri}, where compact expressions up to $m=8$ and $m=9$, respectively, have been reported, after identifying highly
non-trivial commutators identities arising when $m \ge 4$. 

In this work we show that it is indeed possible to get directly rather compact expressions for $\Phi_m$ in terms of right-nested
commutators without much computational effort, sometimes with fewer terms than when expressed in the classical Hall basis. 
This reduction is still more remarkable if the existing commutator identities are introduced at each degree. In addition, the procedure
can be easily extended to the BCH series involving any number of variables,
\begin{equation}  \label{bch-n}
  \exp(X_1) \, \exp(X_2) \cdots \, \exp(X_n) = \exp \big( \Phi(X_1, X_2, \ldots, X_n) \big).
\end{equation}
This can be achieved by considering, instead of the Dynkin presentation (\ref{eq.1.2}), another explicit expression of $\Phi_m$ as a linear
combination of products of $m$ operators $X$ and $Y$ ordered according with the group of permutations. It turns out that such a formula
was also originally obtained by Dynkin and published in his somehow unnoticed 
paper \cite{dynkin49otr}\footnote{We are grateful to Prof. M. M\"uger for bringing  this
reference  to our attention in his ``Notes on the theorem of Baker--Campbell--Hausdorff--Dynkin", available at \newline
\url{https://www.math.ru.nl/~mueger/PDF/BCHD.pdf}}.

One could also consider, of course, left-nested commutators instead and the same results would be still valid with a factor $(-1)^k$, if $k$
is the number of nested commutators.

\section{The BCH series in terms of permutations}

We consider the general case (\ref{bch-n}), i.e.,
\begin{equation}  \label{bchg1}
  \Phi(X_1, X_2, \ldots, X_n) = \sum_{m \ge 1} \Phi_m(X_1, \ldots, X_n),
\end{equation}
where $\Phi_m(X_1, \ldots, X_n)$ is a homogeneous polynomial of degree $m$ in the non-com\-mu\-ta\-ti\-ve 
variables $X_1, \ldots, X_n$. 

Let us denote  by $\varphi_n(X_1, X_2, \ldots, X_n)$ the multilinear part of $\Phi_n(X_1, \ldots, X_n)$, i.e., 
the part obtained by replacing $X_i^2$ by $0$ for all $i$ in
$\Phi_n(X_1, \ldots, X_n)$. Then one has the following remarkable result (see e.g. \cite{loday94sdh}):
\begin{proposition}  \label{propo.1}
  It holds that
\begin{equation}   \label{ei.4}
   \Phi_m(X_1, \ldots, X_n) = \sum_{i_1 + \cdots + i_n = m \atop i_j \ge 0} 
   \frac{1}{i_1! \cdots i_n!} \, \varphi_m(\underbrace{X_1, \ldots, X_1}_{i_1}, \ldots, 
   \underbrace{X_n, \ldots, X_n}_{i_n})
\end{equation}
\end{proposition}
As far as we know, the first proof of Proposition \ref{propo.1} is due to Dynkin \cite{dynkin49otr}. Later on, it was shown that the notion of
Eulerian idempotent leads to a shorter proof \cite{loday94sdh,burgunder08eia}.
The explicit expression of $\varphi_n(X_1, X_2, \ldots, X_n)$
 can be obtained as follows \cite{loday94sdh}. 
Since we are only interested in the
multilinear part of $\Phi_n$, we have to replace  
\[
   \exp(X_1) \, \exp(X_2) \cdots \, \exp(X_n)
\]
by 
\[
    (1+X_1) (1+ X_2)  \cdots (1+ X_n)
\]    
and analyze $\log\big((1+X_1) (1+ X_2)  \cdots (1+ X_n)\big)$, or more specifically, its multilinear part. In other words, we have
to deal with $\log(1+Z)$, where
\[
  Z = \sum_i X_i + \sum_{i < j} X_i X_j + \sum_{i < j < k} X_i X_j X_k + \cdots X_1 X_2 \cdots X_n.
\]  
It is then clear that $\varphi_n$ is of the form
\begin{equation}  \label{egs}
   \varphi_n(X_1, X_2, \ldots, X_n) = \sum_{\sigma \in S_n} c_{\sigma} \, X_{\sigma(1)} X_{\sigma(2)} \cdots X_{\sigma(n)}
\end{equation}
where the sum is extended over all permutations $\sigma$ of $\{1, 2, \ldots, n \}$. The coefficients $c_{\sigma}$ in (\ref{egs}) can be obtained by
analyzing the contribution coming from each power $Z^k$ in the expansion $\log(1+Z) = Z - \frac{Z^2}{2} + \cdots + \frac{(-1)^{k-1}}{k}
Z^k + \cdots$. The situation is similar to the computation of the explicit expression for the Magnus expansion as carried out, in
particular, in \cite{strichartz87tcb}: it turns out that
\[
   c_{\sigma} =  \sum_{i_1 + \cdots + i_m = n  \atop \sigma \in S(i_1, \ldots, i_m)} \frac{(-1)^{m-1}}{m},
\]
where the sum is extended over all ordered partitions $i_1 + \cdots + i_m = n$ of $n$ such that $\sigma \in S(i_1, \ldots, i_m)$,
with
\[
   S(i_1, \ldots, i_m) = \{ \sigma \in S_n \, | \, \sigma(j) < \sigma(j+1) \, \mbox{ for all } \, j \ne i_1 + \cdots + i_{\ell}, \;\; \ell = 1, \ldots, m-1 \}
\]
and $S_n$ denotes the permutation group.
A straightforward computation shows that this number is  $\binom{n - d_{\sigma} - 1}{m-1-d_{\sigma}}$, where $d_{\sigma}$ is the number of
\emph{descents} in $\sigma$. We recall that $\sigma \in S_n$ has an ascent in $i$ if $\sigma(i) < \sigma(i+1)$, $i=1,\ldots, n-1$
and it has a descent in $i$ if $\sigma(i) > \sigma(i+1)$.

In this way one arrives at \cite{strichartz87tcb,loday94sdh}
\[
  c_{\sigma} = \sum_{m = d_{\sigma} + 1}^n \frac{(-1)^{m-1}}{m} \binom{n - d_{\sigma} - 1}{m-1-d_{\sigma}} =
  \frac{(-1)^{d_{\sigma}}}{n}  \, \frac{1}{ \binom{n-1}{d_{\sigma}}}
\]
and finally 
\begin{equation}   \label{ei.1}
  \varphi_n(X_1, X_2, \ldots X_n) = \frac{1}{n}    \sum_{\sigma \in S_n} \, (-1)^{d_{\sigma}} \, \frac{1}{ \binom{n-1}{d_{\sigma}}} \, 
    X_{\sigma(1)} X_{\sigma(2)} \cdots X_{\sigma(n)}.
\end{equation}
At this point some remarks are in order:
\begin{itemize}
 \item The existing relationship between the multilinear part $\varphi_n$ with the Eulerian idempotent can be both ways: either to compute the coefficients $c_{\sigma}$ 
 in (\ref{egs}) by applying
 different descriptions of this object  \cite{solomon68otp,loday89osl} or by providing an explicit combinatorial expression
 for this Eulerian idempotent with (\ref{ei.1}) that allows in particular to characterize its symmetries \cite{burgunder08eia}.
 \item Goldberg \cite{goldberg56tfp} analyzed the formal power series (\ref{bch-n}) when $n=2$ characterizing  the coefficient of the
 general term $X_1^{s_1} X_2^{s_2} \cdots$ in terms of certain polynomials. This result was generalized to an arbitrary $n$ in
 \cite{kobayashi98gta} (see also \cite{helmstetter89sdh}). It turns out that the coefficients $c_{\sigma}$ and the
 explicit expression (\ref{ei.1}) reproduce these previous results.
\end{itemize}

If we restrict ourselves to the case of two variables $X_1 \equiv X$, $X_2 \equiv Y$, then eq. (\ref{ei.4}) reads
\begin{equation}  \label{ei.5}
  \Phi_m(X,Y) = \sum_{i+j = m \atop i,j \ge 1} \, \frac{1}{i!} \frac{1}{j!} \, \varphi_m(\underbrace{X, \ldots, X}_{i}, \underbrace{Y, \ldots, Y}_{j}), 
\end{equation}  
and, according with eq. (\ref{ei.1}), one gets for the first terms
\begin{equation}  \label{fterms}
\aligned
 & \Phi_2(X,Y) = \varphi_2(X, Y) =  \frac{1}{2} X Y - \frac{1}{2} Y X\\
  & \Phi_3(X,Y) = \frac{1}{2} \, \varphi_3(X, X, Y) +  \frac{1}{2} \, \varphi_3(X, Y, Y) \\
  &  \qquad \qquad = \frac{1}{12} XXY - \frac{1}{6} XYX + \frac{1}{12} XYY + \frac{1}{12} YXX - \frac{1}{6} YXY + \frac{1}{12} YYX \\
  & \Phi_4(X,Y) = \frac{1}{6} \, \varphi_4(X,X,X,Y)  + \frac{1}{4} \, \varphi_4(X,X,Y,Y) + \frac{1}{6} \, \varphi_4(X,Y,Y,Y) \\
  &  \qquad \qquad = \frac{1}{24} XXYY - \frac{1}{12} XYXY + \frac{1}{12} YXYX - \frac{1}{24} YYXX
\endaligned
\end{equation}

\section{The BCH series in terms of commutators}

Application of Dynkin--Specht--Wever (DSW) theorem to (\ref{ei.1}) allows one to express $\varphi_n(X_1,\ldots, X_n)$ in terms
of commutators and get an alternative expression for the homogeneous Lie polynomial $\Phi_n$. If we define the
Lie bracketing from right to left by the unique linear map $r$ such that for any word $w = a_1 a_2 \ldots a_{n-1} a_n$ of length $n$
one has $r(w) = [a_1,[a_2,\ldots, [a_{n-1}, a_n] \cdots]]$, the DSW theorem states that 
for each homogeneous Lie polynomial $P$ of degree $n$, it is true that $r(P) = n P$ \cite{reutenauer93fla}. In consequence,
\begin{equation}   \label{ei.1b}
  \varphi_n(X_1, X_2, \ldots, X_n) = \frac{1}{n^2}    \sum_{\sigma \in S_n} \, (-1)^{d_{\sigma}} \, \frac{1}{ \binom{n-1}{d_{\sigma}}} \, 
    [X_{\sigma(1)}, [X_{\sigma(2)} \cdots [X_{\sigma(n-1)},X_{\sigma(n)}] \cdots ]].
\end{equation}

Notice that, as is the case with (\ref{eq.1.2}), not all the commutators (Lie brackets) appearing in 
(\ref{ei.1b}) are linearly independent
among each other, due to antisimmetry and the Jacobi identity. Thus, if one aims to get
a expression in terms of independent commutators, then a particular basis of the vector subspace  
spanned by those commutators in which each generator appears exactly once has to be considered. If we denote this subspace by
$L_n(X_1,\ldots, X_n)$, then $\dim L_n = (n-1)!$.

Among the possible bases of $L_n$, 
the class considered by Dragt \& Forest \cite{dragt83con} is particularly appropriate. 
In forming such a basis, 
one uses only those right-nested brackets ending with a particular but otherwise arbitrary variable selected from the
 collection $X_1, X_2, \ldots, X_n$.  If this
variable is chosen as $X_n$, then the basis is formed by the right-nested brackets of the form  
  \[
      [X_k, [X_{j},\ldots [X_i, X_n] \ldots ]],
  \]
  where the indices $k, j, \ldots i$ are all possible permutations of $\{1, 2, \ldots n-1 \}$. Of course, there are $n$ different such bases, depending on the particular
ending operator one selects. What makes this class of bases specially compelling is the following property. Suppose we have an expression in terms of
products of $n$ distinct operators $X_1, \ldots, X_n$ which is
known to be written as a linear combination of right-nested commutators in $L_n(X_1,\ldots, X_n)$. This is the case, in particular, of $\varphi_n$. Suppose
all the right-nested commutators ending with, say, $X_n$ are used as a basis. Then, in this linear combination, the
coefficient of the right-nested commutator
  $ [X_k, [X_{j},\ldots [X_i, X_n] \ldots ]]$ is \emph{precisely} the coefficient of the permutation $\alpha = (k j \ldots i n)$
in the original expression.

 In consequence, if apply this observation to eq. (\ref{ei.1}), we end up with 
\begin{equation}  \label{ei.2}
  \varphi_n(X_1, X_2, \ldots, X_n) = \frac{1}{n}    \sum_{\sigma \in S_{n-1}} \, (-1)^{d_{\sigma}} \, \frac{1}{ \binom{n-1}{d_{\sigma}}} \, 
    [X_{\sigma(1)}, [X_{\sigma(2)}, \cdots , [X_{\sigma(n-1)}, X_n] \cdots ]],
\end{equation}
just involving the $(n-1)!$ permutations of $S_{n-1}$. Of course, 
similar formulas can be obtained for $\varphi_n$ if instead of choosing $X_n$ as the last element in the right-nested commutator of the
basis one takes any other element $X_j$, $j=1,\ldots, n-1$. In any case, a more compact expression than (\ref{ei.1b}) for $\varphi_n$   
is obtained in this way, since all the commutators are now independent.

\section{Reducing the number of commutators}

Applying formula (\ref{ei.2}) and taking into account eq. (\ref{ei.5}) to the case of two variables we get 
\begin{equation}  \label{bchg1}
  \Phi_m(X,Y)  = \sum_{i+j = m \atop i,j \ge 1} \, \frac{1}{i!} \frac{1}{j!} \, \frac{1}{m} 
    \sum_{\sigma \in S_{m-1}} \,  \frac{(-1)^{d_{\sigma}} }{ \binom{m-1}{d_{\sigma}}} \, 
    [X_{\sigma(1)}, [X_{\sigma(2)}, \cdots , [X_{\sigma(m-1)}, Y] \cdots ]],  \nonumber
\end{equation}
where $X_{\sigma(i)}$, $i=1,\ldots,m-1$, can be either $X$ or $Y$ according with the particular permutation $\sigma$ considered. In particular, for the first
terms we have
 \begin{align}  \label{ft1}
  & \Phi_2(X,Y) = \frac{1}{2} [X,Y] \nonumber \\
  & \Phi_3(X,Y) = \frac{1}{12}  [X,[X,Y]] + \frac{1}{6} [X,[Y,Y]] - \frac{1}{12} [Y,[X,Y]] \\
  & \Phi_4(X,Y) = \frac{1}{24} [X,[X,[Y,Y]]] - \frac{1}{24} [X,[Y,[X,Y]]] + \frac{1}{36} [X,[Y,[Y,Y]]] - \nonumber  \\
   &  \qquad \qquad \;\; - \frac{1}{36} [Y,[X,[Y,Y]]]  \nonumber
\end{align}
If we take into account, however, the obvious property $[z,z] =0$,  then  we get the correct formula for $\Phi_m(x,y)$ up to $m=4$ and a much reduced number of terms in
$\Phi_m$ for $m \ge 5$ than formula (\ref{eq.1.3}) with (\ref{eq.1.2}). 
To substantiate this claim, we have elaborated the code presented in Appendix \ref{apenA}  for the computation of $\varphi_m$
and the functions $\Phi_m(X,Y)$. The number of terms in $\Phi_m$ produced by this code
 is collected in the fifth line of Table
\ref{tab1} up to $m=10$. It is labelled as ``no identities'' to emphasize the fact that no existing identities among commutators have been yet implemented. We include for
comparison the corresponding number of terms of $\Phi_m(X,Y)$ in the classical Hall (third line) and Lyndon bases (fourth line) as obtained by applying the procedure
of \cite{casas09aea}. For completeness, we also write the dimension of each homogeneous subspace $\mathcal{L}_m(X,Y)$ of the free Lie algebra  $\mathcal{L}(X,Y)$.

\begin{table}  
\begin{center}
\begin{tabular}{|c|ccccccccc|}  \hline
$m$  &    2  &  3  &  4  &  5  &  6  &  7  &  8  &  9  &  10   \\ \hline\hline
$\dim \mathcal{L}_m(X,Y)$  &    1  &  2  &  3  &  6  &  9  &  18  &  30  &  56  &  99   \\  \hline
\# terms Hall basis  &   1  &  2  &  1  &  6  &  6  &  18  &  24  &  56  &  86  \\
\# terms Lyndon  basis &    1  &  2  &  1  &  6  &  5  &  18  &  17  &  55  &  55   \\  \hline
no identities  &   1  &  2  &  1  &  8  &  7  &  32  &  31  & 96   &  97  \\  
Grade 4 &   1  &  2  &  1  &  6  &  5  &  24  &  23  &  78  &  78  \\ 
Grade 6  &  1  &  2  &  1  &  6  &  4  &  18  &  17  &  67  &  65 \\  
Compact  &  1  &  2  &  1  &  6  &  4  &  18  &  13  &  38 &   52 \\  \hline
Symmetric compact  & 0  &  2  & 0  &  6  & 0  &  18  & 0  & 42 &   0 \\  \hline

\end{tabular}
\end{center}
\caption{Number of terms in the homogeneous Lie polynomial $\Phi_m(X,Y)$ in the BCH series for the first values of $m$, together with the dimension of $\mathcal{L}_m(X,Y)$.
The last line refers to the symmetric BCH series, eq. (\ref{sy.4}).}
 \label{tab1}
\end{table}  

If we incorporate the identity
\[
     [Y,[X,[X,Y]]] = [X,[Y,[X,Y]]] 
\]
appearing at $m=4$ into the procedure, we get the numbers collected in the line labelled ``Grade 4''. It is remarkable that this number agrees with the corresponding to
the Lyndon basis up to $m=6$, whereas $\Phi_8$ and even $\Phi_{10}$ contain a smaller number of terms than in the classical Hall basis.

Whereas no further identities exist at $m=5$, the following three appear at $m=6$, namely \cite{oteo91tbc}
\begin{eqnarray*}
 \mathrm{(i)} &  & [X,X,X,Y,X,Y] - 2 \, [X,Y,X,X,X,Y] + [Y,X,X,X,X,Y]  = 0, \\
 \mathrm{(ii)} & & [X,X,Y,Y,X,Y] + 3 \, [Y,X,X,Y,X,Y]   - 3 \, [X,Y,X,Y,X,Y]  \\
  & & \;  \qquad\qquad -  [Y,Y,X,X,X,Y] = 0, \\
 \mathrm{(iii)} & & [Y,Y,X,Y,X,Y]  - 2 \, [Y,X,Y,Y,X,Y]  + [X,Y,Y,Y,X,Y] = 0,
\end{eqnarray*}
where we have denoted  $ [X,X,X,Y,X,Y]  :=  [X,[X,[X,[Y,[X,Y]]]]]$, etc., for simplicity.
By incorporating them into the algorithm we get a further reduction, as the line labelled ``Grade 6'' in Table \ref{tab1} clearly shows.

As a matter of fact, a systematic procedure to generate all the existing identities  at a given $m$ can be designed by
using tools of liner algebra, and in particular Gaussian elimination, as is explained in Appendix \ref{apenB} (see also \cite{rencangli95rto}). 
The algorithm can also be used to construct bases of the homogeneous subspace
$\mathcal{L}_m(X,Y)$ for any $m \ge 1$ formed by right-nested commutators in a quite straightforward manner. For completeness, we have collected all such
existing identities up to $m=10$ in the reference \cite{explicitidentities}.

Once the identities have been obtained, to get compact expressions a particular basis has to be identified at each $m$ so that the number of 
vanishing coefficients of $\Phi_m(X,Y)$ is as large as possible. This can be done either by inspection (for small $m$) or applying the technique proposed in 
\cite{kolsrud93mri}. By proceeding in this way, we have been able to get rather compact expressions for
$\Phi_m$, as shown in Table \ref{tab1} (line labelled ``Compact"). The corresponding explicit expressions can be found at \cite{explicitidentities}.
In the reduction process a relevant role is played by the existing symmetries, namely
\[
\begin{aligned}
  &  \Phi_m(-X,-Y) = (-1)^m \, \Phi_m(X,Y) \\
  & \Phi_m(X,Y) = (-1)^{m+1} \, \Phi_m(Y,X)
\end{aligned}
\]   

\section{Further considerations}

The above procedure can also be easily generalized to any number of variables. 
In particular, from eqs. (\ref{ei.4}) and (\ref{ei.2})  we get for the case of three variables
\begin{equation}   \label{sy.3}
\begin{aligned}
  & \Phi_2(x_1,x_2,x_3) = \frac{1}{2} [x_1,x_2] + \frac{1}{2} [x_1,x_3] + \frac{1}{2} [x_2,x_3] \\
  & \Phi_3(x_1,x_2,x_3) = \frac{1}{12} [x_1,[x_1,x_2]] + \frac{1}{12} [x_1,[x_1,x_3]] + \frac{1}{3} [x_1,[x_2,x_3]] - \frac{1}{12} [x_2,[x_1,x_2]]  \\
  & \quad - \frac{1}{6} [x_2,[x_1,x_3]] + \frac{1}{12} [x_2,[x_2,x_3]] - \frac{1}{12} [x_3,[x_1,x_3]] - \frac{1}{12} [x_3,[x_2,x_3]], 
\end{aligned}
\end{equation}
etc. This can be applied to get directly the so-called symmetric BCH formula,
\begin{equation}  \label{sy.4}
  \exp(\frac{1}{2} X) \, \exp(Y) \,  \exp(\frac{1}{2} X) = \exp(\Psi(X,Y)) = \sum_{m=1}^{\infty} \Psi_m(X,Y)
\end{equation}
of great relevance in the design of time-symmetric splitting and composition methods (see e.g. \cite{blanes16aci}, \cite{hairer06gni} and references therein). 
In this case it is easy to show that in general $\Psi_{m}(X,Y) = 0$ when $m$ is even.

Of course, to get compact expressions we have to apply the same procedure as before to the corresponding formulas (\ref{sy.3}) with the obvious replacements
$x_1 \rightarrow X/2$, $x_2 \rightarrow Y$, $x_3 \rightarrow X/2$. It is more advantageous, however, to start with a different expression for (\ref{sy.4}) involving less terms
before applying the reduction procedure. This can be achieved by connecting $\Phi(X,Y)$ and $\exp(\Psi(X,Y))$ as follows:
\[
  \e^{\Psi(X,Y)} = \e^{-\frac{X}{2}} \, \e^X \, \e^{Y} \, \e^{\frac{X}{2}} = \e^{-\frac{X}{2}} \, \e^{\Phi(X,Y)}  \, \e^{\frac{X}{2}}, 
\]
so that 
\begin{equation} \label{sy.5}
   \Psi(X,Y) = \e^{-\mathrm{ad}_{\frac{X}{2}}} \, \Phi(X,Y) = \sum_{k=0}^{\infty} \frac{(-1)^k}{2^k k!} \, \mathrm{ad}_X^k \Phi(X,Y)
\end{equation}
where 
\[
    \mathrm{ad}_A B = [A,B], \qquad \mathrm{ad}_A^j B = [A, \mathrm{ad}_A^{j-1} B], \qquad    \mathrm{ad}_A^0 B = B.  
\]
If we use the compact expressions for $\Phi_m(X,Y)$ obtained in the previous section, then the number of terms in the corresponding $\Psi_m$ determined according with
(\ref{sy.5}) diminishes considerably. For instance, when $m=9$ we get  52 terms instead of 121. By applying the existing identities, this number is further reduced to 42. For
comparison, $\Psi_9$ in the Hall basis contains 56 terms \cite{casas09aea}

The last line in Table \ref{tab1} contains the number of terms of $\Psi_m(X,Y)$  up to $m=9$, whereas the explicit expressions can also be found at \cite{explicitidentities}.

On the other hand, formula (\ref{bchg1}) for $\Phi_m(X,Y)$ can be in fact obtained when the Magnus expansion is used to construct the formal solution of the 
differential equation
\begin{equation}  \label{eqdif}
Y^{\prime}(t)=A(t)Y(t),\qquad\qquad Y(0)=I,
\end{equation}
when $A(t)$ is defined as
\[
   A(t) = \theta(t-1) X + \big( \theta(t) + \theta(t-1) \big) Y,
\]
$\theta(t)$ being the step function. As is well known, the solution of (\ref{eqdif}) can be written as
\[
   Y(t,0) = \exp \Omega(t,0),
\]
where $\Omega$ is an infinite series    
   \begin{equation}
\Omega(t,0)=\sum_{m=1}^{\infty}\Omega_{m}(t,0), \qquad \mbox{ with } \qquad \Omega_m(0,0) = 0,  \label{ME}%
\end{equation}
whose terms are increasingly complex expressions involving time-ordered integrals  of nested commutators of $A$ evaluated at different times. 
An explicit expression for $\Omega_m(t,0)$, $m \ge 1$, in terms of iterated integrals of linear combinations of independent commutators has been obtained
in \cite{arnal18agf}, namely
\begin{equation}   \label{me.com3}
\begin{aligned}
  & \Omega_m(t,0) = \frac{1}{m} \sum_{\sigma \in S_{m-1}} \, (-1)^{d_b} \frac{1}{\binom{m-1}{d_b}} \, 
    \int_0^t dt_1 \int_0^{t_1} dt_2 \cdots \int_0^{t_{m-1}} dt_m \,  \\
    &  \qquad\qquad\qquad  [A(t_{\sigma(1)}), [A(t_{\sigma(2)}) \cdots 
    [A(t_{\sigma(m-1)}), A(t_m)] \cdots ]]
\end{aligned}    
\end{equation}
where $\sigma$ and $d_{\sigma}$ have the same meaning as in (\ref{ei.2}). Notice that, since $Y(t=2,0) = \e^X \e^Y$, then $\log ( \e^X \e^Y ) = \Omega(2,0)$,
and $\Omega_m(2,0)$ as given by (\ref{me.com3}) reproduces exactly the expression of $\Phi_m(X,Y)$ given by (\ref{bchg1}). This can be checked order by order.

\subsection*{Acknowledgements}
FC would like to thank the Isaac Newton Institute for Mathematical Sciences for support and hospitality during the programme ``Geometry,
compatibility and structure preservation in computational differential equations", when work
on this paper was undertaken. This work was supported by EPSRC Grant Number EP/R014604/1 and by 
Ministerio de Econom{\'i}a y Com\-pe\-ti\-ti\-vi\-dad (Spain) through project MTM2016-77660-P (AEI/FEDER, UE).

\appendix
\section{Appendix}  \label{apenA}

The following code in \emph{Mathematica} implements (perhaps not in the most efficient way) formula (\ref{ei.4}):
{\small
\begin{alltt}
fact[n_,m_]:=Product[k[j]!,\{j,1,n-1\}](m-Sum[k[j],\{j,1,n-1\}])!;
table[\{h_\,_\,_\}]:= Table[h];
intable[n_,m_]:= Join[Table[\{X[j], \{r[j],1, If[j == n,
   m-Sum[k[j], \{j,1,n-1\}], k[j]]\}\}, \{j,1,n\}]];
count[n_,m_]:= Table[\{k[j],0,If[j == 1, m, 
   m-Sum[k[i],\{i,1,j-1\}]]\}, \{j,1,n-1\}];
Phinm[f_][n_,m_][\{counter_\,_\,_\}]:= Sum[1/fact[n,m] f[Flatten[
   Join[Table[table[intable[n,m][[j]]],\{j,1,n\}]],2]], counter];
Phi[f_][m_][\{h_\,_\,_\}]:= Phinm[f][Length[\{h\}],m][count[Length[\{h\}],
     m]]/. Table[X[j]-> \{h\}[[j]], \{j,1,Length[\{h\}]\}];
\end{alltt}
}
Next we compute $\Phi_m(X_1, \ldots, X_n)$ with the explicit expression (\ref{ei.1}) for $\varphi_m $ in terms of non-commutative products \texttt{Mm}. This is contained in
\texttt{PhiMm[m][\{$X_1, \ldots, X_n$\}]}: 

{\small
\begin{alltt}  
s[n_]:= Permutations[Range[1, n]];
coef[p_]:= coef[p] = Module[\{b, n\}, b = 0; n = Length[p]; 
   Do[If[p[[i]] > p[[i + 1]], b = b + 1], \{i, 1, n - 1\}];
   (-1)^b/(n Binomial[n - 1, b])];	
ei[\{h_\,_\,_\}]:= Sum[coef[s[Length[\{h\}]][[j]]] Apply[Mm,
   Permutations[\{h\}][[j]]],   \{j,1,Length[\{h\}]!\}];
Ei[\{h_\,_\,_\}]:= ei[Table[x[i], \{i,1,Length[\{h\}]\}]]/. 
     Table[x[j]-> \{h\}[[j]], \{j,1,Length[\{h\}]\}];
PhiMm[m_][\{h_\,_\,_\}]:= PhiMm[m][\{h\}]= Phi[Ei][m][\{h\}];
\end{alltt}
}
Then we express $\varphi_n(X_1, \ldots, X_n)$ in terms of commutators, equation (\ref{ei.2}), and finally $\Phi_m$ by computing \texttt{PhiCmt[m][\{$X_1, \ldots, X_n$\}]}: 
{\small
\begin{alltt}  
Cmt[x_,x_]:= 0;
Cmt[a_\,_\,_,0,b_\,_\,_]:= 0;
Cmt[a_\,_\,_,x_+y_,b_\,_\,_]:= Cmt[a,x,b] + Cmt[a,x,b];
Cmt[a_\,_\,_,n_ x_Cmt,b_\,_\,_]:= n Cmt[a,x,b];
Cmt[a_\,_\,_,t^n_ x_,b_\,_\,_] := t^n Cmt[a,x,b];
Cmt[a_\,_\,_,n_(x_+y_),b_\,_\,_] := Cmt[a,n x,b] + Cmt[a,n y,b];
(* Cmt /: Format[Cmt[x_, y_]]:= SequenceForm["[",x,",",y,"]"]; *)

cmtt[\{b_,a_\}]:=Cmt[b,a];
cmtt[\{a_,h_\,_\,_\}]:=Cmt[a,cmtt[\{h\}]];
basisn[n_]:= basisn[n]= Partition[Flatten[Tuples[
   \{Permutations[Range[1, n-1]], \{n\}\}]], n];
Varphi[\{h_\,_\,_\}]:= Sum[coef[basisn[Length[\{h\}]][[j]]]
   cmtt[Map[x, basisn[Length[\{h\}]][[j]]]]/. Table[x[j]-> 
   \{h\}[[j]], \{j, 1, Length[\{h\}]\}], \{j,1,(Length[\{h\}]-1)!\}];
PhiCmt[m_][\{h_\,_\,_\}]:= PhiCmt[m][\{h\}]= Phi[Varphi][m][\{h\}];
\end{alltt}
}
The first block defines the commutator (just the linearity property and the antisymmetry) with the correct format for output if necessary. Linearity properties in an analogous way should be implemented for the non commutative product \texttt{Mm}. The second block defines the basis and the generic term
$\varphi_n(X_1,\ldots, X_n)$.
Finally, let us remark that since the number of variables is free, the same code allows one to compute both the BCH  \texttt{PhiCmt[m][\{$X,Y$\}]} and the 
symmetric BCH series, \texttt{PhiCmt[m][\{$\frac{1}{2} X,Y, \frac{1}{2} X$\}]}.

\section{Appendix}  \label{apenB}
The algorithm we have applied to generate the identities among commutators and a basis of the homogeneous subspace $\mathcal{L}_m(X,Y)$ formed by
right-nested commutators is a generalization of a procedure proposed in \cite{rencangli95rto}, and 
 can be summarized as follows:

\

\noindent
For each $j=2, \ldots, m$, do
\begin{enumerate}
  \item Generate all possible right-nested commutators $C_i$ involving $j$ operators $X$ and $m-j$ operators $Y$.
 For example, with $m=4$, 
 $$\mathcal{B}_ 4=\{ [X,[X,[X,Y]]], [Y,[X,[X,Y]]], [X,[Y,[X,Y]]], [Y,[Y,[X,Y]]]\}$$
  
  \item Generate the corresponding element $\langle C_i \rangle$ in the homogeneous subspace $\mathcal{U}_m(X,Y)$ of the universal enveloping 
  algebra associated with $\mathcal{L}(X,Y)$. This is done by expanding each commutator $[A,B] = A B - B A$. {For example, on the previous list, for  $C_1=[X,[X,[X,Y]]]$,
  $$<C_1>=XXXY-3XXYX+3XYXX-YXXX$$}
  \item The element $\langle C_i \rangle$ is then a linear combination of words $X_{i_1} X_{i_2} \cdots X_{i_{m}}$, where $X_{i_j}$ is either $X$ of $Y$.
  The total number of words is $\binom{m}{j}$. Once all these words are arranged in a prescribed order, the element commutator $C_i$ can be
  identified with the vector $(a_1, a_2, \ldots, a_p)$ formed by the linear combination. {Then $C_1$ would be $C_1\equiv (1,-3,0,3,0,0,-1,0,0,0,0,0)$.}
  
  \item { Define a matrix $A$ whose rows are formed by these coefficient vectors and augment it to the right with the $m \times m$ identity matrix, forming an block matrix $\left(A | I\right)$. Apply Gauss-Jordan elimination and get the block matrix $\left(M | P\right)$. For $m=4$ we have,
  	
  	\footnotesize{
$$\left(A | I\right)=\left(
\begin{array}{cccccccccccc|cccc}
1 & -3 & 0 & 3 & 0 & 0 & -1 & 0 & 0 & 0 & 0 & 0 & 1 & 0 & 0 & 0 \\
0 & 0 & -1 & 0 & 2 & 0 & 0 & -2 & 0 & 1 & 0 & 0 & 0 & 1 & 0 & 0 \\
0 & 0 & -1 & 0 & 2 & 0 & 0 & -2 & 0 & 1 & 0 & 0 & 0 & 0 & 1 & 0 \\
0 & 0 & 0 & 0 & 0 & 1 & 0 & 0 & -3 & 0 & 3 & -1 & 0 & 0 & 0 & 1 \\
\end{array}
\right)  
$$}
\normalsize{and}
\footnotesize{$$\left(M | P\right)=\left(
\begin{array}{cccccccccccc|cccc}
	1 & -3 & 0 & 3 & 0 & 0 & -1 & 0 & 0 & 0 & 0 & 0 & 1 & 0 & 0 & 0 \\
	0 & 0 & 1 & 0 & -2 & 0 & 0 & 2 & 0 & -1 & 0 & 0 & 0 & 0 & -1 & 0 \\
	0 & 0 & 0 & 0 & 0 & 1 & 0 & 0 & -3 & 0 & 3 & -1 & 0 & 0 & 0 & 1 \\
	0 & 0 & 0 & 0 & 0 & 0 & 0 & 0 & 0 & 0 & 0 & 0 & 0 & 1 & -1 & 0 \\
\end{array}
\right)
$$
}
}
  \item Now the non-vanishing rows on $M$ give the commutators of the basis. The identities we want to find out are obtained after making equal to zero the linear combinations on $P\cdot\mathcal{B}_ n$ corresponding to the vanishing rows.  In our example, since there is one null row on $M$ we get one Grade 4 identity, we equal to zero the last element on the product $P\cdot\mathcal{B}_ 4$, that is, 

$$0=[Y,[X,[X,Y]]]-[X,[Y,[X,Y]]].
$$

\end{enumerate}

\bibliographystyle{siam}

\begin{thebibliography}{10}

\bibitem{explicitidentities}
  \texttt{http://www.gicas.uji.es/Research/bch.html}.

\bibitem{arnal18agf}
{\sc A.~Arnal, F.~Casas, and C.~Chiralt}, {\em A general formula
  for the {M}agnus expansion in terms of iterated integrals of right-nested
  commutators}, J. Phys. Commun., 2 (2018), p.~035024.

\bibitem{blanes16aci}
{\sc S.~Blanes and F.~Casas}, {\em A {C}oncise {I}ntroduction to {G}eometric
  {N}umerical {I}ntegration}, {CRC} Press, 2016.

\bibitem{bonfiglioli10aov}
{\sc A.~Bonfiglioli}, {\em An {ODE}'s version of the formula of {B}aker,
  {C}ampbell, {D}ynkin and {H}ausdorff and the construction of {L}ie groups
  with prescribed {L}ie algebra}, Mediterr. J. Math., 7 (2010), pp.~387--414.

\bibitem{bonfiglioli12tin}
{\sc A.~Bonfiglioli and R.~Fulci}, {\em Topics in {N}oncommutative {A}lgebra.
  {T}he {T}heorem of {C}ampbell, {B}aker, {H}ausdorff and {D}ynkin}, vol.~2034
  of Lecture Notes in Mathematics, Springer, 2012.

\bibitem{burgunder08eia}
{\sc E.~Burgunder}, {\em Eulerian idempotent and {K}ashiwara--{V}ergne
  conjecture}, Ann. Inst. Fourier, 58 (2008), pp.~1153--1184.

\bibitem{casas09aea}
{\sc F.~Casas and A.~Murua}, {\em An efficient algorithm for computing the
  {B}aker--{C}ampbell--{H}ausdorff series and some of its applications}, J.
  Math. Phys., 50 (2009), p.~033513.

\bibitem{dragt83con}
{\sc A.~Dragt and E.~Forest}, {\em Computation of nonlinear behavior of
  {H}amiltonian systems using {L}ie algebraic methods}, J. Math. Phys., 24
  (1983), pp.~2734--2744.

\bibitem{dynkin47eot}
{\sc E.~Dynkin}, {\em Evaluation of the coefficients of the
  {C}ampbell--{H}ausdorff formula}, Dokl. Akad. Nauk. SSSR, 57 (1947),
  pp.~323--326.

\bibitem{dynkin49otr}
{\sc E.~Dynkin}, {\em On the
  representation by means of commutators of the series $\log(e^x e^y)$ for
  noncommutative $x$ and $y$}, Mat. Sb. (N.S.), 25(67) (1949), pp.~155--162 (in
  {R}ussian).

\bibitem{dynkin00spo}
{\sc E.~Dynkin}, {\em Calculation of the
  coefficients in the {C}ampbell--{H}ausdorff series}, in Selected {P}apers of
  {E.B.} {D}ynkin with {C}ommentary, E.~Dynkin, A.~Yushkevich, G.~Seitz, and
  A.~Onishchik, eds., {A}merican {M}athematical {S}ociety, 2000, pp.~31--35.

\bibitem{goldberg56tfp}
{\sc K.~Goldberg}, {\em The formal power series for $\log(\e^x \e^y)$}, Duke
  Math. J., 23 (1956), pp.~13--21.

\bibitem{hairer06gni}
{\sc E.~Hairer, C.~Lubich, and G.~Wanner}, {\em Geometric {N}umerical
  {I}ntegration. {S}tructure-{P}reserving {A}lgorithms for {O}rdinary
  {D}ifferential {E}quations}, Springer-Verlag, {S}econd~ed., 2006.

\bibitem{helmstetter89sdh}
{\sc J.~Helmstetter}, {\em S\'erie de {H}ausdorff, d'une alg\`ebre de {L}ie et
  projections canoniques de l'alg\`ebre enveloppante}, J. Algebra, 120 (1989),
  pp.~170--199.

\bibitem{kobayashi98gta}
{\sc H.~Kobayashi, N.~Hatano, and M.~Suzuki}, {\em Goldberg's theorem and the
  {B}aker--{C}ampbell--{H}ausdorff formula}, Physica A, 250 (1998),
  pp.~535--548.

\bibitem{kolsrud93mri}
{\sc M.~Kolsrud}, {\em Maximal reductions in the {B}aker-{H}ausdorff formula},
  J. Math.\ Phys., 34 (1993), pp.~270--285.

\bibitem{rencangli95rto}
{\sc R.-C. Li}, {\em Raising the order of unconventional schemes for ordinary
  differential equations}, PhD thesis, {U}niversity of {C}alifornia at
  {B}erkeley, 1995.

\bibitem{loday89osl}
{\sc J.-L. Loday}, {\em Op\'erations sur l'homologie cyclique des alg\`ebres
  commutatives}, Invent. Math., 96 (1989), pp.~205--230.

\bibitem{loday94sdh}
 {\sc J.-L. Loday}, {\em S\'erie de
  {H}ausdorff, idempotents {E}ul\'eriens et alg\`ebres de {H}opf}, Expo. Math.,
  12 (1994), pp.~165--178.

\bibitem{mclachlan19tla}
{\sc R.~McLachlan and A.~Murua}, {\em The {L}ie algebra of classical
  mechanics}, tech. rep., arXiv:1905.07554, 2019.

\bibitem{oteo91tbc}
{\sc J.~Oteo}, {\em {T}he {B}aker-{C}ampbell-{H}ausdorff formula and nested
  commutator identities}, J. Math.\ Phys., 32 (1991), pp.~419--424.

\bibitem{reutenauer93fla}
{\sc C.~Reutenauer}, {\em Free {L}ie {A}lgebras}, vol.~7, Oxford University
  Press, 1993.

\bibitem{solomon68otp}
{\sc L.~Solomon}, {\em On the {P}oincar\'e--{B}irkhoff--{W}itt theorem}, J.
  Comb. Theory, 4 (1968), pp.~363--375.

\bibitem{sornborger99hom}
{\sc A.~Sornborger and E.~Stewart}, {\em Higher-order methods for simulations
  on quantum computers}, Phys. Rev. A, 60 (1999), pp.~1956--1965.

\bibitem{strichartz87tcb}
{\sc R.~S. Strichartz}, {\em The {C}ampbell--{B}aker--{H}ausdorff--{D}ynkin
  formula and solutions of differential equations}, J. Funct. Anal., 72 (1987),
  pp.~320--345.

\bibitem{varadarajan84lgl}
{\sc V.~Varadarajan}, {\em Lie {G}roups, {L}ie {A}lgebras, and {T}heir
  {R}epresentations}, Springer-Verlag, 1984.

\bibitem{wilcox67eoa}
{\sc R.~Wilcox}, {\em Exponential operators and parameter differentiation in
  quantum physics}, J. Math. Phys., 8 (1967), pp.~962--982.

\end{thebibliography}

\end{document}